\documentclass[conference]{IEEEtran}
\IEEEoverridecommandlockouts
\usepackage{cite}
\usepackage{amsmath,amssymb,amsfonts}
\usepackage{algorithmic}
\usepackage{graphicx}
\usepackage{textcomp}
\usepackage{xcolor}
\usepackage{url}
\usepackage{booktabs}
\usepackage[hidelinks]{hyperref}
\makeatletter
\def\@IEEEsectpunct{}
\let\SLTorigthebibliography\thebibliography
\renewcommand{\thebibliography}[1]{%
  \SLTorigthebibliography{#1}%
  \fontsize{9}{10.5}\selectfont
  \setlength{\itemsep}{2pt plus 0.5pt}%
}
\makeatother
\def\BibTeX{{\rm B\kern-.05em{\sc i\kern-.025em b}\kern-.08em
    T\kern-.1667em\lower.7ex\hbox{E}\kern-.125emX}}
\begin{document}

\bstctlcite{IEEE:BSTcontrol}

\title{StreamTN: A Low-Latency Streaming Chinese Text Normalization Model for Streaming TTS in Dialogue Systems}

\author{
\IEEEauthorblockN{
Wenhao Li\textsuperscript{1},
Jinrui Liang\textsuperscript{1},
Haoyu Zhang\textsuperscript{1},
Jingbin Hu\textsuperscript{1},
Xiaming Ren\textsuperscript{1},
Hanke Xie\textsuperscript{1},
Huakang Chen\textsuperscript{1},\\
Chengyou Wang\textsuperscript{1},
Dake Guo\textsuperscript{1},
Linhan Ma\textsuperscript{1},
Su Feng\textsuperscript{2},
Houdun Liu\textsuperscript{2},
Yunxiang Chen\textsuperscript{2},
Lei Xie\textsuperscript{1,*}
}
\IEEEauthorblockA{
\textsuperscript{1}Audio, Speech and Language Processing Group (ASLP@NPU),\\
Northwestern Polytechnical University, Xi'an, China\\
\textsuperscript{2}Shenzhen Pimei Technology Co., Ltd., Shenzhen, China\\
\href{mailto:wenhao.li@mail.nwpu.edu.cn}{wenhao.li@mail.nwpu.edu.cn}, \href{mailto:lxie@nwpu.edu.cn}{lxie@nwpu.edu.cn}}
\thanks{* indicates the corresponding author.}
}

\maketitle
\vspace{-2em}

\begin{abstract}
Text-to-Speech (TTS) is an essential module that provides spoken responses in a spoken dialogue system (SDS) centered on a large language model (LLM). To ensure accurate TTS synthesis, responses generated by an LLM must be converted into TTS-readable formats via a Text Normalization (TN) module, imposing strict low-latency requirements in real-time SDS scenarios. Existing TN solutions are largely rule-based, rely on manual engineering, and generalize poorly to unseen patterns. Although an LLM itself can perform TN through prompt engineering, it faces key limitations: high first-token latency due to non-streaming processing, hallucination risks, and degraded intelligence or reasoning when the core LLM module is fine-tuned solely for TN. To address these challenges, we propose \textit{StreamTN}, a lightweight LLM-based Chinese streaming TN model. Built on Qwen3-0.6B, StreamTN employs a dual-track streaming framework in which input tokens and output tokens are processed on two parallel tracks, enabling low-latency real-time inference without complex prompting. Moreover, task-specific fine-tuning yields superior TN performance and fewer hallucinations than rule-based systems and general-purpose LLMs. We also introduce a TN benchmark that spans diverse text scenarios, providing a comprehensive evaluation standard for speech generation in spoken dialogue systems. Experiments demonstrate the effectiveness of StreamTN in accuracy and inference latency.
\end{abstract}


\begin{figure*}[htbp]
  \centering
  \includegraphics[width=0.64\textwidth]{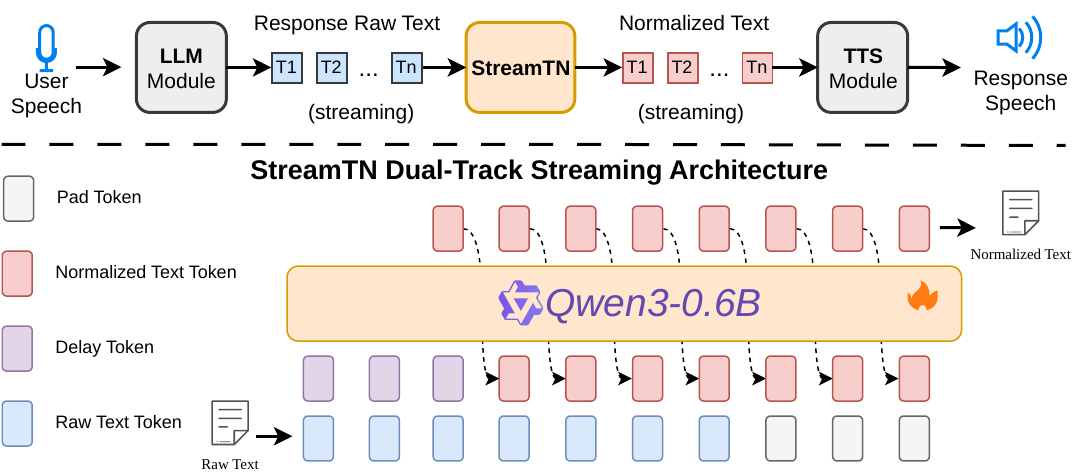}
  \vspace{-1mm}
  \caption{The architecture of our proposed StreamTn model.} 
  \vspace{-2mm}
  \label{fig:model_architecture} 
\end{figure*}

\section{Introduction}

Spoken dialogue systems (SDS) have become an important interface for natural human--computer interaction, especially with the rapid development of large language models (LLMs). Modern LLM-centered SDS typically follow either an end-to-end paradigm, where speech understanding, response generation, and speech synthesis are jointly modeled, or a cascaded paradigm, where Automatic Speech Recognition (ASR), an LLM module, and Text-to-Speech (TTS) are organized as independent components~\cite{wen2017network,young2013pomdp,zhang2020recent,WavChat,DBLP:conf/acl/CuiYJMZWGK25}. Recent speech-language models further explore speech-native or real-time interaction with LLMs~\cite{SpeechGPT,Moshi,Mini-Omni,LLaMA-Omni}. Although end-to-end systems have shown promising progress, cascaded systems remain widely used in practical applications due to their controllability, modularity, ease of debugging, and flexibility in integrating external knowledge or retrieval modules.

In a cascaded SDS, the textual response produced by the LLM module serves as the direct input to the TTS module. However, LLM-generated responses often contain non-standard words (NSWs), such as numbers, dates, time expressions, phone numbers, units, chemical formulas, and mathematical expressions. These forms are readable to humans but can be ambiguous or unsuitable for direct speech synthesis. For example, a single number may correspond to different pronunciations depending on whether it appears in a date, a phone number, an amount, or a mathematical expression. Without proper text normalization (TN), such ambiguity may cause mispronunciations, unnatural prosody, or even synthesis failures in downstream TTS systems.

Text normalization has been extensively studied as the task of converting non-standard written expressions into context-appropriate verbalizations, with TTS being one of its primary applications. Traditional industrial TN systems often rely on handcrafted rules, regular expressions, and WFST-based grammars~\cite{gorman2016pynini,ebden2015kestrel,nemo_text_processing}, which provide strong controllability but require substantial language-specific engineering. To reduce reliance on manually designed rules, neural approaches formulate TN as a sequence modeling task, predicting context-appropriate verbalizations from written input~\cite{zhang2019neural}. However, purely neural approaches may produce rare but severe unrecoverable errors, which is particularly problematic for TTS. For Mandarin TN, hybrid approaches have been proposed to combine rule-based systems with neural models such as multi-head self-attention~\cite{zhang2020hybrid}. More recently, FlatTN introduces a rule-guided flat-lattice Transformer to integrate expert rules into end-to-end Chinese TN and releases a large-scale Chinese TN dataset~\cite{flattn}. These studies significantly advance TN modeling and resource construction, but they mainly focus on offline or sentence-level normalization rather than low-latency streaming normalization in LLM-centered dialogue systems.

Recent progress in LLMs has further motivated prompt-based TN. PolyNorm, for example, explores few-shot LLM-based TN for TTS and introduces a multilingual benchmark covering diverse normalization phenomena~\cite{polynorm2025}. This line of work demonstrates the potential of LLMs for scalable TN across languages. Nevertheless, directly applying prompt-based LLMs to cascaded SDS remains challenging because the TN module must process partial LLM outputs and provide normalized text to the TTS module in real time. Prompt-based normalization usually requires waiting for sufficient or complete context, introduces additional latency, and may still suffer from unstable formatting or hallucinated outputs. In contrast, our goal is not only to improve TN accuracy, but also to enable streaming input processing and streaming output generation with controllable first-packet delay.

A key challenge in LLM-centered cascaded SDS is that both the upstream LLM response and the downstream TTS synthesis are expected to operate in a streaming manner, especially as recent TTS systems increasingly emphasize scalable and low-latency streaming synthesis~\cite{cosyVoice2}. In this setting, the TN module is no longer a conventional offline text preprocessing component. Instead, it must incrementally convert partial LLM outputs into TTS-readable text while preserving pronunciation correctness and semantic consistency. Therefore, an effective TN module for dialogue TTS should satisfy two requirements simultaneously: it should generate accurate and TTS-readable normalized text, and it should do so incrementally with low latency.

This new application scenario also exposes a benchmark gap. Although evaluation methods for dialogue systems have been broadly studied~\cite{Survey_on_evaluation_methods_for_dialogue_systems}, existing TN datasets and evaluations mainly focus on conventional NSW categories, offline sentence-level normalization, or multilingual TTS normalization, and they do not fully reflect the distribution, diversity, and latency requirements of real dialogue responses generated by LLMs. In particular, dialogue-oriented TTS must handle not only common expressions such as numbers, dates, time, phone numbers, and units, but also increasingly frequent complex outputs involving scientific terms, chemical formulas, and mathematical expressions. Without a scenario-aligned TN benchmark, it is difficult to systematically compare different methods under the practical requirements of cascaded SDS, including normalization accuracy, robustness to diverse text patterns, and streaming latency.

To address these challenges, we propose StreamTN, a lightweight LLM-based Chinese streaming text normalization model for cascaded SDS. Built on Qwen3-0.6B~\cite{Qwen3TechnicalReport}, StreamTN introduces a dual-track streaming architecture that decouples raw-text input tokens and normalized-text output tokens into two parallel tracks. This design enables the model to begin normalization after receiving only a small number of input tokens, rather than waiting for the complete LLM response. Through task-specific fine-tuning, StreamTN learns structured normalization patterns while avoiding complex prompting and reducing hallucination risks.

In addition, we construct a Chinese dialogue-oriented TN benchmark tailored to streaming TTS in cascaded SDS. The benchmark covers 14 categories, including common NSW types such as numbers, dates, time, phone numbers, and units, as well as challenging scientific expressions such as chemical formulas and mathematical equations. It provides a scenario-aligned evaluation setting for measuring not only normalization accuracy, but also robustness across diverse dialogue outputs and latency under streaming inference.

The main contributions of this work are summarized as follows:
\begin{itemize}
\setlength{\itemsep}{0pt}
\setlength{\parskip}{0pt}
\setlength{\parsep}{0pt}
\setlength{\topsep}{2pt}
\item We propose StreamTN, a dedicated Chinese streaming TN model with a dual-track architecture that enables incremental input processing and output generation with controllable first-packet delay.
\item We construct a dialogue-oriented Chinese TN benchmark covering 14 categories of practical normalization scenarios, filling the gap of TN evaluation for streaming TTS in cascaded dialogue systems.
\item Experiments show that StreamTN provides competitive normalization quality at low first-packet delay and improves further when additional context is available.
\item To facilitate reproducibility and future research, we will release the trained StreamTN model and the accompanying benchmark upon publication.
\end{itemize}

A demonstration is available online.\footnote{\url{https://supernova-neko.github.io/Stream-TN/}}

\section{Method}
\subsection{Overall architecture}

As illustrated in Figure~\ref{fig:model_architecture}, StreamTN is inserted between the upstream LLM module and the downstream TTS module in a cascaded spoken dialogue system. The LLM module first generates a textual response from the user's spoken query, and the TN module converts this response into a TTS-readable form before speech synthesis. In conventional cascaded systems, TN is usually performed by a rule-based module or a lightweight neural model after the complete textual response is available. StreamTN replaces this offline TN component and processes the LLM response incrementally, so that normalized text can be passed to the TTS module before the full response is finished~\cite{DBLP:conf/interspeech/AroraTFJSKT025}.

This setting differs from standard sentence-level text normalization. In an offline TN system, the model can access the whole input sentence and use both left and right context to disambiguate non-standard words. In a streaming dialogue system, however, the input arrives token by token from the LLM module. The TN model therefore needs to decide whether the currently available prefix already contains enough information for normalization. Emitting too early may lead to incorrect readings, while waiting for too much context increases the first-packet latency of the TTS system. Thus, the architecture needs an explicit mechanism to control the amount of delayed context used for generation.

StreamTN is built on Qwen3-0.6B to reuse its general text modeling ability. The original model, however, follows a conventional autoregressive formulation in which input and output tokens are arranged along a single sequence. This form is not well suited to streaming TN, because output generation is coupled with the availability of the input sequence. Inspired by Qwen3-Omni~\cite{Qwen3-Omni} and delayed-stream modeling~\cite{DBLP:journals/corr/abs-2509-08753}, we introduce a dual-track structure into the text model. The raw LLM output and the normalized text are represented on two separate tracks, and their embeddings are fused before being fed into the Transformer backbone.

With this structure, the input track provides the currently available raw-text context, while the output track carries the delayed normalized-text history. The model can therefore generate normalized tokens autoregressively while still consuming new raw tokens from the upstream LLM. When the LLM stream is still active, the model uses both tracks; when the raw input stream has ended, the input track is padded and generation continues based on the output history until the end-of-sequence token is produced. This formulation keeps the TN model compatible with standard autoregressive decoding, while allowing the amount of input lookahead to be controlled by a fixed delay parameter.

The following subsection gives the formal definition of the dual-track streaming process, including the construction of the two tracks, representation fusion, the training objective, and first-packet delay.

\subsection{Dual-track streaming formulation}

Given a raw text sequence produced by the upstream LLM module, we denote the input token sequence as \(\mathbf{x}=(x_1,x_2,\ldots,x_n)\), where \(x_t\) denotes the raw-text token at time step \(t\), and \(n\) is the input length. The corresponding normalized target sequence is denoted as \(\mathbf{y}=(y_1,y_2,\ldots,y_m)\), where \(y_i\) denotes the normalized-text token at output step \(i\), and \(m\) is the target length. A conventional autoregressive text generation model typically learns the conditional distribution \(p_{\theta}(\mathbf{y}\mid\mathbf{x})\), where \(\theta\) denotes the model parameters. In offline text normalization, this formulation assumes that the complete input sequence is available before decoding begins. In the streaming setting considered here, however, the model only observes a prefix of the LLM response at each decoding step, making it necessary to reformulate the normalization process under partial input context.

Let \(d\) denote the token-delay parameter, which controls how many raw input tokens are observed before the first normalized token is emitted. For each aligned step \(t\), we use \(\bar{x}_t\) and \(\bar{y}_t\) to denote the symbolic states on the raw-input track and the output-history track, respectively. The raw-input track is defined as
\begin{equation}
\bar{x}_t =
\begin{cases}
x_t, & 1 \le t \le n, \\
\langle \mathrm{pad} \rangle, & t > n,
\end{cases}
\end{equation}
and the output-history track is delayed by \(d\) aligned steps:
\begin{equation}
\bar{y}_t =
\begin{cases}
\langle \mathrm{delay} \rangle, & 1 \le t \le d, \\
y_{t-d}, & d < t \le m+d, \\
\langle \mathrm{pad} \rangle, & t > m+d.
\end{cases}
\end{equation}
Here, \(\langle \mathrm{delay} \rangle\) and \(\langle \mathrm{pad} \rangle\) are conceptual alignment symbols rather than learned vocabulary tokens. They have different semantic roles: the former indicates that output generation has not started, whereas the latter indicates that a track has ended. In the implementation, both symbols are realized as zero vectors after embedding. Let \(\Phi(\cdot)\) denote the track-to-embedding mapping based on the token-embedding table \(E(\cdot)\) inherited from Qwen3-0.6B:
\begin{equation}
\Phi(u)=E(u), \qquad
\Phi(\langle \mathrm{delay} \rangle)
=
\Phi(\langle \mathrm{pad} \rangle)
=
\mathbf{0},
\end{equation}
where \(u\) is an ordinary raw or normalized token. The two tracks share the same embedding space and are fused by element-wise addition:
\begin{equation}
\mathbf{z}_t = \Phi(\bar{x}_t) + \Phi(\bar{y}_t).
\end{equation}
Let \(T=\max(n,m+d)\) denote the aligned length. During training, the last fused position is removed so that every normalized target is predicted from its preceding causal state. The resulting sequence is processed by the Qwen3 Transformer with a causal attention mask:
\begin{equation}
\mathbf{h}_{1:T-1}
=
\mathrm{Qwen3}_{\theta}
\left(
\mathbf{z}_{1:T-1};\mathbf{M}_{\mathrm{causal}}
\right),
\end{equation}
where positional information is applied internally by Qwen3 according to the aligned position indices. Under this causal shift, the hidden state at aligned step \(i+d-1\) contains the raw prefix \(x_{\le \min(n,i+d-1)}\) and the normalized history \(y_{<i}\), but not the current target \(y_i\). The next-token distribution is therefore
\begin{equation}
p_{\theta}
\left(
y_i \mid x_{\le \min(n,i+d-1)},y_{<i}
\right)
=
\mathrm{softmax}
\left(
W_o \mathbf{h}_{i+d-1} + \mathbf{b}_o
\right),
\end{equation}
where \(W_o\) and \(\mathbf{b}_o\) are the output projection matrix and bias, respectively.

This alignment gives the streaming factorization
\begin{equation}
\begin{aligned}
p_{\theta}(\mathbf{y}\mid\mathbf{x})
=
\prod_{i=1}^{m}
p_{\theta}
\left(
y_i \mid x_{\le \min(n,i+d-1)},y_{<i}
\right).
\end{aligned}
\end{equation}
The delay parameter \(d\) determines the trade-off between latency and available input context. In particular, \(y_1\) is predicted from \(\mathbf{h}_d\) after the first \(d\) raw tokens have been observed. At the next aligned step, the model consumes the embedding of the previously generated token together with the next available raw-token embedding and predicts \(y_2\). Thus, no target token is included in the hidden state used to predict itself.

During training, supervision is applied only to valid normalized target tokens; zero-padded positions are excluded by the target mask. The training objective is the negative log-likelihood
\begin{equation}
\begin{aligned}
\mathcal{L}_{\mathrm{TN}}
=
-\sum_{i=1}^{m}
\log
p_{\theta}
\bigl(
y_i
\mid
x_{\le \min(n,i+d-1)}, y_{<i}
\bigr).
\end{aligned}
\end{equation}
This objective trains the model to perform normalization under partial input context, rather than relying on the complete raw sequence.

During inference, StreamTN follows the same alignment using the autoregressive key--value cache. The model first receives \(d\) raw tokens and predicts \(\hat{y}_1\) from \(\mathbf{h}_d\). For \(i>1\), aligned step \(t=i+d-1\) combines the next raw-token embedding, when available, with \(E(\hat{y}_{i-1})\), and the resulting hidden state predicts \(\hat{y}_i\). After the raw stream ends, the raw-input state becomes \(\langle \mathrm{pad} \rangle\), whose mapped representation is \(\mathbf{0}\), and decoding continues until an end-of-sequence token is produced or a predefined maximum generation budget is reached.

The implementation measures the model-side first-token computation time after the first \(d\) raw tokens have become available. In a complete spoken dialogue system, the end-to-end first-packet delay additionally includes the time required by the upstream LLM to produce these tokens:
\begin{equation}
\label{eq:fpd_e2e}
\mathrm{FPD}_{\mathrm{e2e}}(d)
=
T^{\mathrm{LLM}}_{\mathrm{wait}}(d)
+
T^{\mathrm{TN}}_{\mathrm{first}}(d).
\end{equation}
Here, \(T^{\mathrm{LLM}}_{\mathrm{wait}}(d)\) denotes the upstream waiting time, whereas \(T^{\mathrm{TN}}_{\mathrm{first}}(d)\) includes embedding the available prefix, the initial Transformer prefill, and computation of the first normalized-token logits. Reporting these two terms separately distinguishes intrinsic TN computation latency from the token-arrival rate of a particular upstream LLM.

\subsection{Dataset construction}
\subsubsection{Taxonomy Design and Category Consolidation}\mbox{}\par\nopagebreak
Prior work, including FlatTN~\cite{flattn}, organizes Chinese text normalization categories according to the Non-Standard Word (NSW) standard, which adopts a fine-grained classification scheme. However, modern LLM-based training requires a more concise data representation consisting of paired raw text and normalized text to support efficient manual inspection and model optimization~\cite{zhang2026instruction}. Building on the taxonomy of FlatTN~\cite{flattn}, the original categories are consolidated into ten unified types to simplify the label space while preserving coverage. In addition, to address user queries involving scientific content and the corresponding complex outputs generated by the LLM module, four new categories are introduced: simple chemical substances, complex chemical equations, simple mathematical formulas, and complex mathematical equations. Based on this refined taxonomy, an evaluation benchmark is constructed for Chinese SDS~\cite{kiela2021dynabench,huang2023c}.

\begin{figure}[t]
  \centering
  \includegraphics[width=\linewidth]{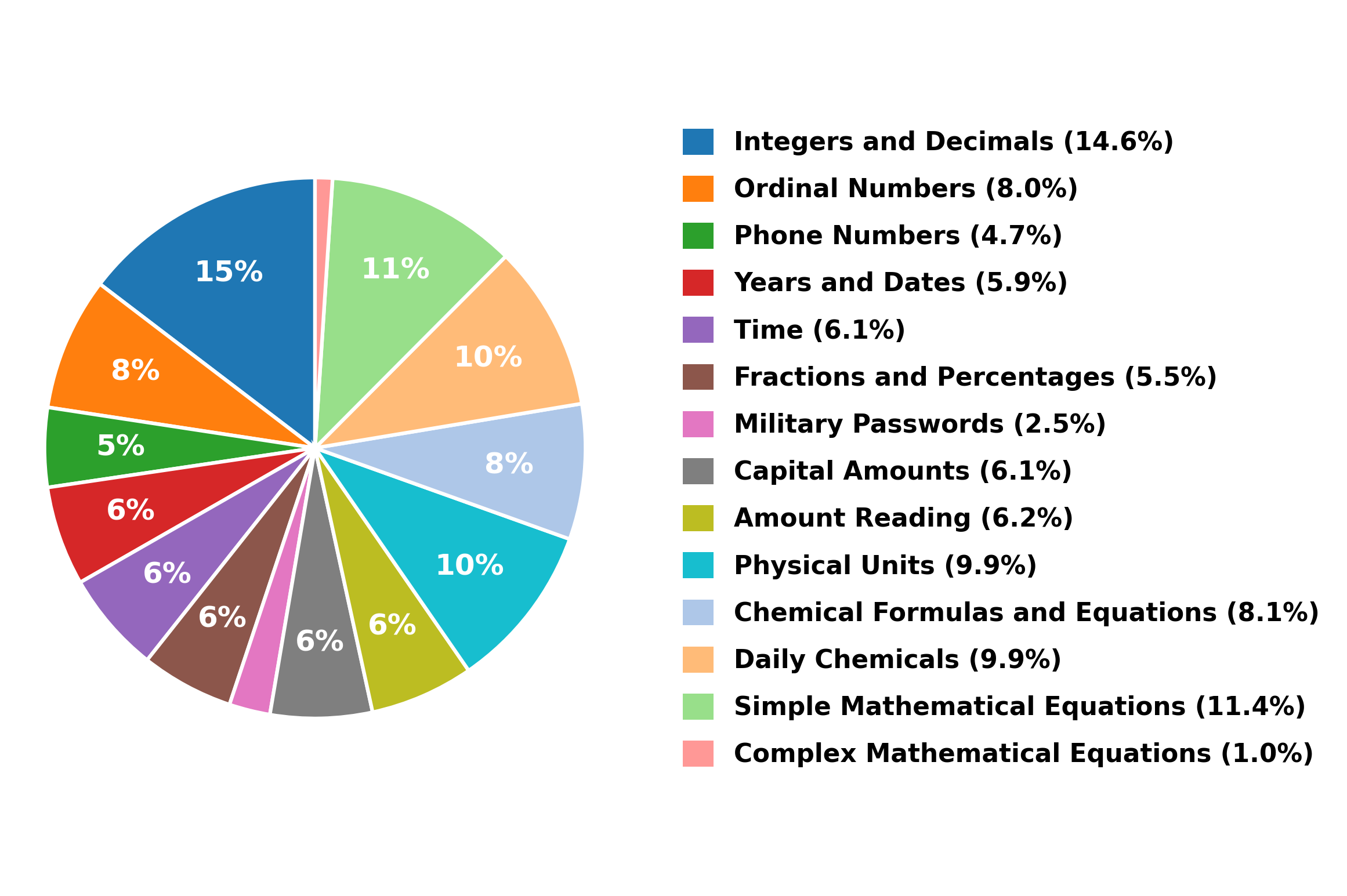}
  \caption{Data distribution of text normalization categories.}
  \label{fig:tn_distribution}
\end{figure}

\subsubsection{Data Collection and Annotation Pipeline}\mbox{}\par\nopagebreak

After consolidating the taxonomy, we defined category-specific TN guidelines based on the pronunciation and readability requirements of Chinese TTS systems. The guidelines specify the expected normalization form for each category and were applied consistently during annotation and evaluation. We constructed the training corpus from two complementary sources. First, we screened publicly available Chinese text from DuReader~\cite{DuReader}, whose questions and supporting documents originate from real-world Baidu Search and Baidu Zhidao data. Regular-expression-based detectors were used to identify text containing normalizable NSW patterns, and only the matched samples were retained as raw inputs.

Second, because naturally occurring data provide limited coverage of the four newly introduced scientific categories---simple chemical substances, complex chemical equations, simple mathematical formulas, and complex mathematical equations---we used Qwen3-32B~\cite{Qwen3TechnicalReport} to generate additional raw-text samples. These AI-generated samples were manually reviewed to remove unnatural expressions and obvious synthetic artifacts. Normalized targets for both the filtered open-source samples and the AI-generated samples were produced using DeepSeek-R1-70B~\cite{guo2025deepseek} under the same predefined TN specifications. To assess annotation quality, we randomly sampled 5\% of the training set for manual verification. Of the inspected samples, 98.2\% were judged consistent with the predefined TN guidelines. In addition, all 1,262 held-out test references were manually inspected and corrected according to the same guidelines.

\section{Experiments}
\subsection{Experimental setup}
\subsubsection{Datasets}\mbox{}\par\nopagebreak
A comprehensive dataset covering diverse TN categories has been constructed, comprising 95,793 training samples and 1,262 held-out test samples. The training set is used for model optimization, whereas the test set evaluates generalization to unseen and challenging TN scenarios.

\subsubsection{Baselines}\mbox{}\par\nopagebreak
To comprehensively evaluate StreamTN across different modeling paradigms, we compare it with four representative baselines:

\begin{itemize}
\setlength{\itemsep}{0pt}
\setlength{\parskip}{0pt}
\setlength{\parsep}{0pt}
\setlength{\topsep}{2pt}
    \item \textbf{WeTextProcessing}~\cite{wetextprocessing}: a widely used open-source rule-based toolkit covering common TN patterns such as numerals, dates, and units. We retain its default rule set.

    \item \textbf{FlatTN}~\cite{flattn}: a rule-guided end-to-end Chinese TN model that integrates expert rules and is evaluated on the same test set.

    \item \textbf{BiLSTM}~\cite{sunkara2021neural,sproat2017rnn}: a fully data-driven lightweight encoder-decoder that learns normalization mappings directly from paired data without handcrafted rules.

    \item \textbf{Qwen3-0.6B}~\cite{Qwen3TechnicalReport}: a general-purpose pre-trained language model evaluated with task-oriented prompting, but without task-specific fine-tuning or streaming adaptation.
\end{itemize}

\subsubsection{Evaluation Metrics}\mbox{}\par\nopagebreak

Model performance is evaluated using character-level micro-averaged precision (Micro-P), recall (Micro-R), and F1 (Micro-F1). For each prediction--reference pair, we first obtain a minimum-cost Levenshtein alignment and count character matches ($M$), substitutions ($S$), deletions ($D$), and insertions ($I$). Following the implementation used in our evaluation, matches are treated as true positives, while substitutions contribute to both false positives and false negatives; insertions contribute to false positives and deletions to false negatives. The alignment counts are accumulated over the entire evaluation set before computing the metrics, rather than averaging sentence-level scores. Accordingly, the three primary metrics are defined as
\begin{equation}
\begin{aligned}
P_{\mathrm{micro}}&=\frac{\sum_i M_i}{\sum_i(M_i+S_i+I_i)},\\
R_{\mathrm{micro}}&=\frac{\sum_i M_i}{\sum_i(M_i+S_i+D_i)}.
\end{aligned}
\end{equation}
\begin{equation}
F_{1,\mathrm{micro}}=\frac{2P_{\mathrm{micro}}R_{\mathrm{micro}}}{P_{\mathrm{micro}}+R_{\mathrm{micro}}}.
\end{equation}
This corpus-level aggregation gives each aligned character equal weight and is used for both the overall comparison and the category-level analysis.

\subsubsection{Experimental Environment}\mbox{}\par\nopagebreak
The model uses PyTorch with Hugging Face Transformers. Unless noted, main experiments use full-parameter fine-tuning. For ablation, we also examine parameter-efficient fine-tuning via LoRA, which adds low-rank adapters to the attention projections (${q\_proj}$, ${k\_proj}$, ${v\_proj}$, ${o\_proj}$) while freezing backbone weights. In LoRA, the rank is 8, the scaling factor is 32, and the dropout is 0.05. Training runs on four NVIDIA A6000 GPUs with dynamic batching. The learning rate starts at $1\times10^{-5}$, decaying to $1\times10^{-6}$ using cosine annealing over 5,000 steps. To account for variability arising from training initialization, each trainable model configuration is independently trained with five random seeds, and its normalization metrics are reported as mean $\pm$ standard deviation. Deterministic rule-based and off-the-shelf baselines are evaluated once. StreamTN inference uses greedy decoding on a single NVIDIA A6000 GPU. For end-to-end first-packet-delay measurements, the upstream Qwen3-32B~\cite{Qwen3TechnicalReport} is served with vLLM using tensor parallelism on two A6000 GPUs and generates approximately 20 tokens/s; its tokens are streamed to StreamTN as soon as they become available. The reported delay follows Eq.~\eqref{eq:fpd_e2e} and includes both the waiting time for the first $d$ upstream tokens and the StreamTN computation required to produce the first normalized token.

\subsection{Model performance}

\subsubsection{Overall Performance}\mbox{}\par\nopagebreak

Table~\ref{tab:overall performance} reports the Micro-F1 and Micro-Precision of all methods. The rule-based WeTextProcessing baseline performs reasonably on predefined patterns, but its limited rule coverage restricts its ability to handle diverse and context-dependent expressions. FlatTN obtains a Micro-F1 of 0.7569, compared with 0.7853 for WeTextProcessing. The general-purpose Qwen3-0.6B baseline performs substantially worse than the task-specific models. Even with task-oriented prompting, its outputs remain susceptible to format instability and hallucination. More elaborate prompts can improve normalization quality, but they also increase prefill cost and first-token latency, making this approach less suitable for real-time dialogue.

At the selected four-frame delay, StreamTN achieves a Micro-F1 of $0.8937\pm0.0009$, which is comparable to the BiLSTM result of $0.8941\pm0.0015$, while obtaining a higher Micro-Precision ($0.8931\pm0.0022$ versus $0.8677\pm0.0012$). This operating point is chosen to balance normalization quality and first-packet delay rather than to maximize offline accuracy. As shown in Table~\ref{tab:different_frames_of_delay}, allowing more input context further improves StreamTN and enables it to surpass BiLSTM in Micro-F1. Overall, StreamTN provides a stronger practical trade-off than the evaluated baselines by combining competitive normalization performance with incremental processing and controllable first-packet latency.

\begin{table}[th]
  \caption{Micro-F1 and Micro-Precision of Different Models}
  \label{tab:overall performance}
  \centering
  \begin{tabular}{lcc}
    \toprule
    \textbf{Method} & \textbf{Micro-F1} & \textbf{Micro-P} \\
    \midrule
    WeTextProcessing~\cite{wetextprocessing} & 0.7853 & 0.7586 \\
    Qwen3-0.6B~\cite{Qwen3TechnicalReport} & 0.4872 & 0.5282 \\
    FlatTN~\cite{flattn} & $0.7569 \pm 0.0019$ & $0.7390 \pm 0.0017$ \\
    BiLSTM~\cite{sunkara2021neural} & $\mathbf{0.8941} \pm 0.0015$ & $0.8677 \pm 0.0012$ \\
    StreamTN & $0.8937 \pm 0.0009$ & $\mathbf{0.8931} \pm 0.0022$ \\
    \bottomrule
  \end{tabular}
\end{table}

\subsubsection{Model Performance Across Different Categories}\mbox{}\par\nopagebreak

Table \ref{tab:tn_types_performance} reports the micro-averaged precision, recall, and F1-score of StreamTN across 14 TN categories. StreamTN performs best on physical units, capital amounts, and time expressions, achieving Micro-F1 scores of 0.977, 0.975, and 0.967, respectively. Strong results are also obtained for integers and decimals, fractions and percentages, and years and dates, all of which exceed 0.95 Micro-F1. These categories generally follow regular and frequently observed normalization patterns. In contrast, complex mathematical equations and chemical formulas remain the most challenging categories, with Micro-F1 scores of 0.750 and 0.791, respectively. Their specialized notation, long verbalizations, and structural diversity increase the difficulty of character-level normalization. Overall, the small standard deviations indicate that StreamTN performs consistently across training seeds.

\begin{table}[th]
  \caption{Micro Precision, Recall, and F1 Scores of Different TN Types}
  \label{tab:tn_types_performance}
  \centering
  \resizebox{\columnwidth}{!}{%
  \begin{tabular}{ l  c  c  c }
    \toprule
    \textbf{TN Type} &
    \textbf{Micro-P} &
    \textbf{Micro-R} &
    \textbf{Micro-F1} \\
    \midrule
    Integers and Decimals    & 0.959 $\pm$ .006 & 0.963 $\pm$ .004 & 0.961 $\pm$ .004 \\
    Ordinal Numbers          & 0.923 $\pm$ .010 & 0.929 $\pm$ .005 & 0.926 $\pm$ .007 \\
    Phone Numbers            & 0.919 $\pm$ .008 & 0.917 $\pm$ .008 & 0.918 $\pm$ .008 \\
    Years and Dates          & 0.949 $\pm$ .009 & 0.957 $\pm$ .009 & 0.953 $\pm$ .009 \\
    Time                     & 0.956 $\pm$ .002 & 0.978 $\pm$ .003 & 0.967 $\pm$ .001 \\
    Fractions and Percentages& 0.964 $\pm$ .009 & 0.947 $\pm$ .006 & 0.955 $\pm$ .007 \\
    Military Passwords       & 0.925 $\pm$ .003 & 0.924 $\pm$ .004 & 0.925 $\pm$ .003 \\
    Capital Amounts          & 0.970 $\pm$ .004 & 0.979 $\pm$ .001 & 0.975 $\pm$ .002 \\
    Amount Reading           & 0.927 $\pm$ .006 & 0.919 $\pm$ .011 & 0.923 $\pm$ .007 \\
    Physical Units           & 0.976 $\pm$ .003 & 0.978 $\pm$ .003 & 0.977 $\pm$ .003 \\
    Chemical Formulas        & 0.785 $\pm$ .008 & 0.797 $\pm$ .006 & 0.791 $\pm$ .006 \\
    Daily Chemicals          & 0.913 $\pm$ .008 & 0.884 $\pm$ .009 & 0.898 $\pm$ .007 \\
    Simple Math Equations    & 0.956 $\pm$ .005 & 0.943 $\pm$ .005 & 0.950 $\pm$ .005 \\
    Complex Math Equations   & 0.749 $\pm$ .014 & 0.751 $\pm$ .007 & 0.750 $\pm$ .009 \\
    \bottomrule
  \end{tabular}%
  }
\end{table}

\subsection{Ablation Study}
\subsubsection{Impact of LoRA Fine-Tuning and System Prompt}\mbox{}\par\nopagebreak
Table \ref{tab:ablation experiment} evaluates LoRA-based parameter-efficient fine-tuning and system-prompt conditioning within the StreamTN framework. The full StreamTN model achieves the best results, with a Micro-F1 of $0.8937\pm0.0009$ and a Micro-Precision of $0.8931\pm0.0022$. Replacing full-parameter fine-tuning with LoRA reduces Micro-F1 to $0.6335\pm0.0012$ and Micro-Precision to $0.6250\pm0.0015$, indicating that updating only low-rank adapters is insufficient to learn the precise transformations required by TN. Adding a system prompt also yields a small but consistent decrease, producing a Micro-F1 of $0.8852\pm0.0018$ and a Micro-Precision of $0.8824\pm0.0021$. Overall, the ablation results demonstrate the importance of full-parameter fine-tuning and show that StreamTN can generate structured normalized text without carefully engineered prompts.

\begin{table}[th]
  \caption{Results of ablation experiment.}
  \label{tab:ablation experiment}
  \centering
  \begin{tabular}{ l  c  c } 
    \toprule
    \textbf{Method} &
    \textbf{Micro-F1} &
    \textbf{Micro-P} \\ 
    \midrule
    StreamTN                & $\mathbf{0.8937} \pm 0.0009$   & $\mathbf{0.8931} \pm 0.0022$ \\ 
    \hspace{2em} w/ LoRA Fine-tuning     & $0.6335 \pm 0.0012$            & $0.6250 \pm 0.0015$ \\
    \hspace{2em} w/ System Prompt     & $0.8852 \pm 0.0018$            & $0.8824 \pm 0.0021$ \\
    \bottomrule
  \end{tabular}
\end{table}

\subsubsection{Impact of Streaming Strategy}\mbox{}\par\nopagebreak
Table \ref{tab:different_frames_of_delay} reports the performance of StreamTN under different delay-frame configurations. The non-streaming model achieves the highest Micro-F1 (0.9639) and Micro-Precision (0.9620) because it has access to the complete input sequence, but it cannot produce normalized output until the entire sequence has been received. Within the streaming framework, increasing the delay allows the model to observe more contextual information before normalization, thereby resolving ambiguities in context-dependent expressions and substantially improving normalization performance. Specifically, as the delay increases from 1 to 16 frames, Micro-F1 rises from 0.7030 to 0.9239 and Micro-Precision from 0.7135 to 0.9303. This improvement, however, is accompanied by an increase in FPD from 75~ms to 756~ms. We therefore adopt the 4-frame configuration in the main experiments as a practical trade-off between normalization quality and responsiveness; it achieves a Micro-F1 of 0.8937 and a Micro-Precision of 0.8931 with an FPD of 213~ms.

\begin{table}[th]
  \caption{StreamTN performance under different delay frames. 
  FPD (ms) denotes First Packet Delay in milliseconds. 
  The configuration marked with $^{*}$ indicates the setting adopted in our main experiments.}
  \label{tab:different_frames_of_delay}
  \centering
  \setlength{\tabcolsep}{2.5pt}
  \resizebox{\linewidth}{!}{
  \begin{tabular}{ l  l  c  c  c}
    \toprule
    \textbf{Strategy} &
    \textbf{Delay frame} &
    \textbf{Micro-F1} &
    \textbf{Micro-P} &
    \textbf{FPD (ms)} \\
    \midrule
    Non-Streaming    & -              & $\textbf{0.9639} \pm 0.0011 $ & $\textbf{0.9620} \pm 0.0012   $ & -   \\
    \midrule
    Streaming        & 1 frame        & $0.7030 \pm 0.0010   $        & $0.7135 \pm 0.0018   $ & 75 \\
                     & 2 frames       & $0.7952 \pm 0.0015   $        & $0.8062 \pm 0.0026   $ & 121 \\
                     & 4 frames$^{*}$ & $0.8937 \pm 0.0009   $        & $0.8931 \pm 0.0022   $ & 213 \\ 
                     & 8 frames       & $0.9072 \pm 0.0011   $        & $0.9093 \pm 0.0020   $ & 394 \\
                     & 16 frames      & $0.9239 \pm 0.0008   $        & $0.9303 \pm 0.0015   $ & 756 \\
    \bottomrule
  \end{tabular}
  }
\end{table}

\section{Conclusions}
This paper presented StreamTN, a Chinese text normalization model designed for low-latency cascaded spoken dialogue systems. StreamTN combines task-specific fine-tuning with a dual-track architecture that processes incoming text and normalized output on separate tracks, allowing normalization to proceed before the complete input is available. On our benchmark, the four-frame configuration achieved a Micro-F1 of 0.8937 with a first-packet delay of 213~ms. Its normalization performance was comparable to the BiLSTM baseline while supporting incremental processing, and additional input context produced further improvements at the cost of higher latency. The category-level results show that most errors occur in structurally complex mathematical and chemical expressions. We will release the model and benchmark to support further work on streaming text normalization. Future work will focus on these difficult categories and extend StreamTN to additional languages and application domains.

\newpage
\bibliographystyle{IEEEtran}
\bibliography{mybib}

\end{document}